\documentclass[aps,prb,twocolumn,superscriptaddress,amsmath,amssymb,footinbib,longbibliography]{revtex4-2}
\usepackage[english]{babel}
\usepackage{graphicx}
\usepackage{dcolumn}
\usepackage{bm}
\usepackage{subfigure}
\usepackage[utf8x]{inputenc}
\usepackage{color}
\usepackage{textcomp}
\usepackage[colorlinks,bookmarks=false,citecolor=blue,linkcolor=red,urlcolor=blue]{hyperref}

\newcommand{\op}[1]{%
    \fontdimen12\textfont3=2pt\fontdimen12\scriptfont3=1.4pt%
    \!\null\mathop{\vphantom{#1}\smash{#1}}\limits_{\sim}\null\!}
\newcommand{\xref}[1]{\protect\ref{#1}}
\newcommand{\figref}[1]{Fig.~\protect\ref{#1}}

\newcommand{\fmref}[1]{(\protect\ref{#1})}
\def\bra#1{\langle \, {#1} \, | \,}
\def\ket#1{\, | \, {#1} \, \rangle}

\renewcommand{\eqref}[1]{Eq.~(\protect\ref{#1})}

\newcommand{\vecops}[1]{\op{\vec{s}}_{#1}}
\newcommand{\ops}[1]{\op{s}_{#1}}
\newcommand{\vecopsprod}[2]{\vecops{#1} \cdot \vecops{#2}}
\newcommand{\Mean}[1]{\big\langle\big\langle \; {#1}\; \big\rangle\big\rangle}

\begin{document}

\title{Approximating thermodynamic equilibrium observables of quasi one-dimensional magnetic molecules and chains by thermal DMRG -- a user perspective}
\author{Lukas Horstmann}
\email{lukas.horstmann@uni-bielefeld.de}
\author{J\"urgen Schnack}
\email{jschnack@uni-bielefeld.de}
\affiliation{Fakult\"at f\"ur Physik, Universit\"at Bielefeld, Postfach 100131, D-33501 Bielefeld, Germany}

\begin{abstract}
Density matrix renormalization group methods or tensor network methods in general 
can not only be used to determine ground states but also
to evaluate thermal equilibrium properties. Although convergence is superior
for one-dimensional quantum spin systems with nearest neighbor exchange 
and open boundary conditions, 
the hope is that these methods can as well be employed to approximate 
magnetic observables of magnetic molecules with more complex interaction patterns. 
Here, we 
assume the position of a user and
study the accuracy that can realistically be achieved for several archetypical 
quasi one-dimensional structures using the TenPy suite. In our study, we aim 
at systems that are too large for exact diagonalization or Krylov space methods.
\end{abstract}

\maketitle

\section{Introduction}
\label{sec-1}

Small quantum spin systems such as small magnetic molecules can
be rationalized by completely diagonalizing their Hamiltonian if
the dimension of the underlying Hilbert space is small enough 
even if the Hamiltonian is complex 
\cite{GaP:GCI93,SZP:JPI96,BCC:JCC99,Wal:PRB00,BSS:JMMM00,ScS:IRPC10,HeS:PRB19,GhK:Mag25,WSU:SciPostPhysCodeb.70,WSU:SciPostPhysCodeb.70-r0.4}.
In such schemes, symmetries may be employed in order to block-structure 
the Hamiltonian matrix.
However, quantum spin systems are modelled in Hilbert spaces that grow exponentially
with the number of spins which renders a complete diagonalization of the Hamiltonian
matrix impossible already for rather small systems \cite{Sch:CP19}.
If symmetries are lost due to e.g.\ anisotropy or exchange patterns 
with low point-group symmetry this happens already for rather small systems 
such as six spins $s=5/2$ to provide an idea.

Approximate methods such as the finite-temperature Lanczos method (FTLM) ease the problem
to some extent since they make quantum spin systems accessible for Hilbert space 
dimensions of up to about $10^{11}$ \cite{JaP:PRB94,PRE:COR17,SRS:PRR20}.
Our hope, however, is that methods based on the density matrix renormalization group theory (DMRG)
or tensor network approaches in general 
\cite{Whi:PRL1992,Whi:PRB93,ExS:PRB03,Sch:RMP05,Jec:PTPS08,SCH:AoP11,UNM:PRB12},
are capable of treating even bigger spin systems as for instance in 
\cite{ExS:PRB03} for ground state properties or in
\cite{UNM:PRB12} for low-lying excitations.
For equilibrium expectation values DMRG schemes have been developed in the early 2000s, 
see \cite{Vid:PRL04,FeW:PRB05,Whi:PRL09}, 
but not applied to large magnetic molecules to the best of our knowledge.
Moreover, an early paper of our group came to the conclusion 
that observables at low temperatures
are only inaccurately reproduced \cite{HRS:JMMM19}.

Meanwhile, new schemes have been developed and made accessible through
open sources. In this investigation, we employ the TenPy suite of which 
we apply two approaches to realize the imaginary time evolution: the so-called
\emph{Variational Apply Evolution}
and the \emph{Time Evolution Block Decimation}, 
see \cite{ZMK:PRB15,PKS:AP19,Mis:arXiv26} and references therein.
We compare our results to the already well established finite-temperature Lanczos method (FTLM),
compare e.g.\ \cite{JaP:PRB94,PRE:COR17,SRS:PRR20},
to assess the applicability of matrix product state (MPS)
approximations for the evaluation of thermal expectation values. 
To this end, we 
assume the position of a user and
select a few typical quasi one-dimensional examples with non-trivial 
interaction patterns for our study.

Our conclusion so far is that good convergence of magnetic observables is
achieved in the case of open spin chains with just nearest neighbor interactions
as expected. The convergence suffers for more spins per unit cell, 
more complex interaction patterns as well as single-ion anisotropy.
However, for most realistic cases the results are sufficiently accurate 
down to experimentally employed temperatures of 1.8~K
and of great help if other options are not available.
In particular in cases
of large systems that cannot be treated by exact diagonalization or 
Krylov space methods such as FTLM
thermal DMRG provides an option free of the sign problem inherent
in Quantum Monte Carlo as demonstrated e.g.\ for an iron chain 
or a spin tube in the following.

The paper is organized as follows. In Section \ref{sec-2} we introduce 
model and methods before we present our results in Sec.~\xref{sec-3}.
The article closes with a discussion in Section~\ref{sec-4}. 
Readers familiar with DMRG and FTLM can directly continue with Sec.~\xref{sec-3}.

\section{Methods}
\label{sec-2}

\subsection{Spin models}

In this paper, we investigate systems that can be simulated by 
spin Hamiltonians such as
\begin{align}
\label{ham}
    \op{H}
    =& \sum_{i\leq j}^{N} 
    \vecops{i}\cdot \mathbf{J}_{ij}\cdot\vecops{j}
+ \mu_B \vec{B}\cdot \sum_{i=1}^{N} g_i \vecops{i}
    \ ,
\end{align}
where $\op{\vec{s}}_i$ denotes the spin vector operator at site $i$. 
A tilde marks operators in general. 
The matrices $\mathbf{J}_{ij}$ contain exchange integrals of isotropic and 
anisotropic type as well as single-ion anisotropy.
With the convention used in \fmref{ham}
antiferromagnetic exchange interactions are given by positive exchange integrals.

In the following examples we will specify the employed Hamiltonian 
more precisely.

\subsection{Thermal DMRG}

Thermal DMRG is a variant of the density matrix renormalization group algorithm
which originally provides an approximation of the ground state. 
This algorithm was expanded over time to improve both the methods and 
their efficiency so that it now provides a larger application space 
than just calculating ground states. 
For thermodynamics the starting point is a pure state at 
$T= \infty$ ($\beta=\frac{1}{k_B T}=0$) which is constructed by adding an 
auxiliary copy to the system.
This state is evolved with imaginary-time evolution in order to cool
the state to the desired temperature $T = T_{\text{min}}$. 
The time evolution is executed within the DMRG algorithm that uses 
matrix-product states (MPS) as an approximation to full 
quantum many-body states.

MPS have the form \cite{Sch:AP11,Mis:arXiv26}
\begin{align}
    \ket{\Phi[A]} = &
    \\
    \sum_{s_1 \dots s_N}^{d} \sum_{\alpha_1 \dots \alpha_{N-1}} 
    &A_{1,\alpha_1}^{s_1} A_{\alpha_1,\alpha_2}^{s_2}
    \dots A_{\alpha_{N-2},\alpha_{N-1}}^{s_{N-1}} A_{a_{N-1},1}^{s_N}
    \nonumber
    \\
    &\ket{s_1, s_2, \dots, s_N}
    \ ,
    \nonumber
\end{align}
where each state is expressed by a train of tensors 
and each tensor represents one physical spin site (or unit cell). 
The value of $d$ is $(2s_i+1)$ and thus might be different if spins
$s_i$ are different. The dimensions $\alpha_i$ are called bond dimensions 
and restricted to a maximum value $m$  according to the impact 
in the DMRG algorithm. 
The impact is related to the contribution to the von Neumann entropy
\begin{align}
S_{A|B} = - \sum_{\alpha} \omega_{\alpha} log_2(\omega_{\alpha}),
\end{align}
with $\omega_{\alpha}$ being the eigenvalues of the reduced density matrix $\rho_A$ 
of a subsystem A of the complete spin system \cite{Neu:NGWG27}. 
Due to their structure matrix product states inherit the area law by construction 
\cite{ECP:RMP10,Mis:arXiv26}. 
The area law states that the entanglement between two subsystems is proportional 
to a value based on the geometry of the total system. 
In the case of one-dimensional gapped systems it is proportional to a constant value 
and as a consequence also allows the bond dimension 
$m$ to be a constant independent of the length of the system \cite{KuS:NC20,Sch:AP11}.
The bond dimension $m$, i.e., the linear dimension of the employed matrices,
defines the dimension of the approximation and serves
as a control parameter of the accuracy. 
A larger $m$ gives higher accuracy, but needs more computation time.
If the investigated quantum system possesses symmetries these can be employed. 
Here we use the $\op{S}^z$-symmetry whenever it is possible to boost performance.

Using this setup, one can update the sites locally as in the original DMRG formalism 
\cite{Whi:PRL1992,Whi:PRB93}. There one constructs a local effective Hamiltonian 
$\op{H}_{\text{eff}}$ on two sites, which minimizes the energy on these sites 
and then moves one site to the left/right and repeats the process over and over until the total energy is minimized \cite{Sch:AP11}. 
This allows a high numerical efficiency
since we do not need to save and optimize the whole system at once 
\cite{Sch:AP11,ZMK:PRB15}.
While this method is efficient, it is unfortunately also 
limited to specific geometric structures, because of the consequences 
of the area law \cite{KuS:NC20}. In comparison, FTLM is limited only
by the dimension of the Hilbert space \cite{SRS:PRR20}.

In order to evaluate thermal equilibrium observables, we employ the purification approach. 
Instead of just taking the density matrix $\rho$ of the system, a copy of the system is used to construct a pure state
\begin{align}
    \ket{\Phi_\rho} = \sum_a \sqrt{p_a} \,   \ket{a_P} \otimes  \ket{a_Q}
\end{align}
from which a density matrix can be created.
This scheme can be used to calculate observables at arbitrary 
temperatures $T$ \cite{Mis:arXiv26}.

To this end, one starts from the trivial pure state $\rho(0)$, 
which represents the density matrix at $\beta = \frac{1}{k_B T} = 0$, i.e., $T=\infty$.
This state is then evolved by means of imaginary-time evolution 
$\op{U}(\beta) = \exp(- \frac{\beta}{2} \op{H} )$ to the 
desired inverse temperature $\beta$. The time evolution operator only acts on the physical part 
and by tracing out the auxiliary party the original density matrix can be recreated.
For the numerical realization of the time evolution different approaches have been developed.
Here we focus on two, both included in TenPy: variational apply of time evolution MPOs, which  
use an approximation of $\op{U}(t)$ as a MPO with compact representation:   
$W^{\text{I}}, W^{\text{II}}$ \cite{ZMK:PRB15} and 
Time Evolution Block Decimation (TEBD), which is very efficient but limited 
by the geometry of the spin system 
\cite{Vid:PRL03,DKS:JSM04}, 
for details compare \cite{PKS:AP19,Mis:arXiv26}.

\subsubsection{Variational Apply Evolution -- thDMRG-VarMPO}

A simple way to evolve the state is by constructing the time evolution operator 
$\op{U}$ as a MPO 
and use the sweep structure of DMRG to update the state $\ket{\Phi}$.
To construct the MPO we use here a modified version of Runge-Kutta 
to the Euler step and construct a $W^{\text{II}}$
as a sufficient approximation to the evolution operator \cite{ZMK:PRB15}. 
To increase the accuracy one can create higher order versions by using a set of smaller time steps;
these versions still have a compact MPO representation. Here we choose the 2nd order 
and use the time steps \cite{ZMK:PRB15}
\begin{align}
    \delta t_1 = \frac{1+i}{2} \delta t  \quad \quad   \&  \quad \quad  \delta t_2 = \frac{1-i}{2} \delta t  
    \ .
\end{align}
Applying now $\op{ \,U}_{\text{$W^{\text{II}}$}}(\delta t_1)$ 
and $\op{ \, U}_{\text{$W^{\text{II}}$}}(\delta t_2)$ to the state in an alternating fashion, 
while performing convergence sweeps, achieves the desired 
imaginary time evolved state $\ket{\Phi(\beta)}$.

\subsubsection{Time Evolution Block Decimation -- thDMRG-TEBD}

Time Evolution Block Decimation (TEBD) is based on the Trotter decomposition of the Hamiltonian 
into independent blocks which commute with each other \cite{Suz:CMP76}. 
Here mostly the second-order decomposition is used due to its efficiency 
in relation between calculation cost and accuracy. 
One splits the Hamiltonian  into local even ($\op{H}_i$, $i$ even) 
and odd ($\op{H}_i$, $i$ odd) parts 
so that each $\op{H}_i$ only acts on the sites
$i$ and $i+1$. Since all even and all odd $\op{H}_i$ commute with each other,
the time-evolution operator can be written as a product
\begin{align}
    \op{U}^{(2)}_{\text{TEBD}}(\delta t) 
    = 
    e^{-i \frac{\delta t}{2} \op{H}_{\text{even}} } e^{-i \delta t \op{H}_{\text{odd}} } 
    e^{-i \frac{\delta t}{2} \op{H}_{\text{even}} }
\end{align}
with a second-order error in $\delta t$ \cite{Vid:PRL04,Sch:AP11}.
Because of the locality of evolution gates in TEBD and the frequent update after each step, the method is very efficient and keeps the bond dimension small.
The downside of the locality is that the method is limited to nearest-neighbour 
interactions. Every longer-range interaction must be expressed as a 
chain of nearest-neighbour gates with grouping of site or swap-gates \cite{Cat:PRB25}, 
details on the later will not be further discussed here .

\subsubsection{Non-trivial models in DMRG}

The best use case for DMRG are spin chains with open boundary conditions 
that have a gapped groundstate and only nearest neighbour interactions \cite{KuS:NC20,Sch:AP11}. 
For this class of systems the bond dimension can be kept small, and we are able to increase 
the length $L$ of the chain to large values. 
Beyond this we have to consider some modifications to make other classes of models 
possible or more feasible for TEBD. We will discuss this in the following.

\begin{figure}[ht!]
\centering
\includegraphics*[width=0.80\columnwidth]{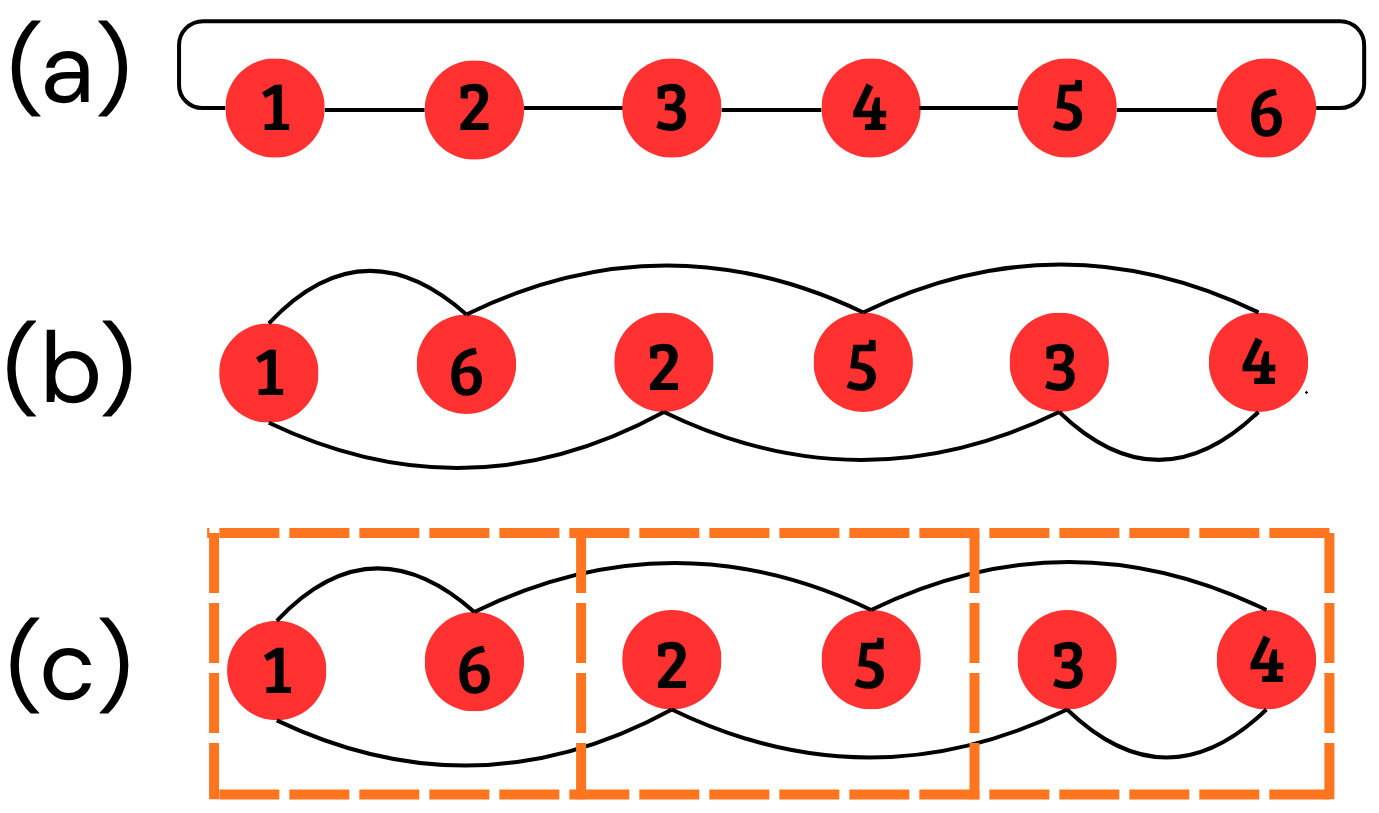}
\caption{Schematic view on how to improve periodic boundary condition: (a) MPS chain with one long interaction
(b) reordering of the sites to only next-nearest interaction without changing physics
(c) grouping two neighbouring sites together so the interaction is only nearest neighbour}
\label{thDMRGfMM-f-1}
\end{figure}

Periodic boundary conditions are a general problem of the MPS formalism. 
The concept of a chain being periodic cannot be implemented directly in a one-dimensional MPS chain.
To simulate a periodic boundary condition one adds a long range 
interaction between the first and the last site.
This long range interaction increases the size of all MPOs and has effects on all
sites, see \figref{thDMRGfMM-f-1}(a).
This not only denies the usage of TEBD, but also slows down convergence and efficiency 
of other time-evolution methods, and it increases calculation times per sweep \cite{Sch:AP11}.
In order to address this problem one has to adjust the geometry of the chain 
so it is suitable for one-dimensional MPS. 
For DMRG it does not matter in which order the sites are sorted as long 
as the specific interactions to each site remain the same \cite{Sch:AP11}.
Therefore, to avoid the long-range interaction one reorders the sites 
in a way that the longest interaction range is now of length two 
instead of length $N$, compare \figref{thDMRGfMM-f-1}(b).

TEBD still does not work since it is limited to nearest-neighbour interaction.
This problem can be addressed by grouping two sites, see \figref{thDMRGfMM-f-1}(c).
This creates $N/2$ sites where each site now has the dimension of 
$\text{dim}(\mathcal{H}(s_i)) \cdot \text{dim}(\mathcal{H}(s_{i+1}))$, hence the Hilbert space dimension of both previous spins combined. This solves the problem, but increasing the dimension makes calculation times larger  compared to two single spins. Therefore, the recommendation is 
to check for other available options before grouping sites \cite{WYH:ACIE}.

\subsubsection{Further technical details of DMRG calculations}

In order to achieve accurate calculations without spending unnecessary calculation times, 
an adaptive adjustment of parameters is done.  At  higher temperatures the state can be described 
by smaller matrices, so it is reasonable to enlarge the bond dimension $m$ 
during the process \cite{Sch:AP11}.

To find a good balance between accuracy and run time we use the truncation error 
as control parameter. If after a cooling step the truncation error exceeds a threshold, 
the bond dimension is increased, here mostly by a fixed amount of $10-50$ 
based on the size of the problem and the results of first test meassurements.
To avoid unreasonable or uncontrollable growth we also set a maximum bond dimension 
$m_{\text{max}}$. The bond dimension given in the following examples 
has to be understood as $m_{\text{max}}$.

For most of our calculations a truncation error 
threshold of $10^{-9}$ to $10^{-7}$ was used, 
which is larger compared to usual ground state applications
and constitutes a good compromise to keep the bond dimension 
manageable.

We would also like to mention, that we did not aim at a systematic study of
the achievable accuracy as function of bond dimension $m$ and imaginary-time
step $d\beta$, but more on realistic computational conditions regarding memory and time.
All examples were run on a dual processor machine with two AMD EPIC processors
with in total 256 hyper-threaded cores and 2 TB RAM in total.

\subsection{FTLM}

The finite-temperature Lanczos method (FTLM) approximates thermal equilibrium observables
as \cite{JaP:PRB94,PRE:COR17}
\begin{align}
\label{ftlm}
\Mean{\op{O}}
&=
\frac{\sum_r \bra{r}e^{-\beta \op{H}/2} \op{O} e^{-\beta \op{H}/2}\ket{r}}
{\sum_r \bra{r}e^{-\beta \op{H}/2} e^{-\beta \op{H}/2}\ket{r}}
   \ .
\end{align}
The method has been investigated by several authors, see e.g.\ 
\cite{ADE:PRB03,SSR:PRB18,SRH:ZNA20,MoT:PRR20};
it can be discussed from various points of view:
(a) The expectation value with respect to a random vector $\ket{r}$
constitutes an approximation of the trace. The action of $e^{-\beta \op{H}/2}$ 
is then evaluated by a Lanczos method or a Chebychev approach \cite{SGS:ZNA21}.
(b) One can also consider the random vector $\ket{r}$ a representative of 
infinite temperature ($\beta=0$) and the evaluation of the action of 
$e^{-\beta \op{H}/2}$ an imaginary-time evolution.
The average over random vectors improves the accuracy \cite{SRS:PRR20}.

\section{Results}
\label{sec-3}

\subsection{A delta chain close to the critical point}

We start with this example since it was critically discussed in 
Ref.~\cite{HRS:JMMM19}.

\begin{figure}[ht!]
\centering
\includegraphics*[width=0.90\columnwidth]{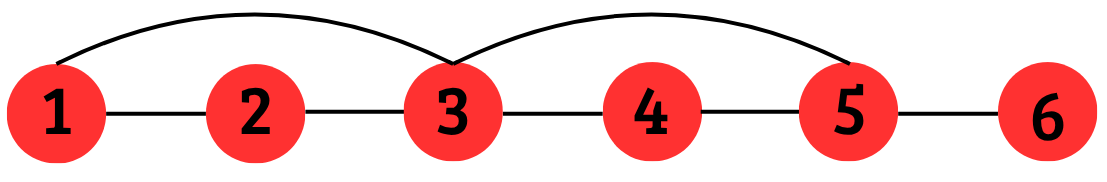}
\includegraphics*[width=0.90\columnwidth]{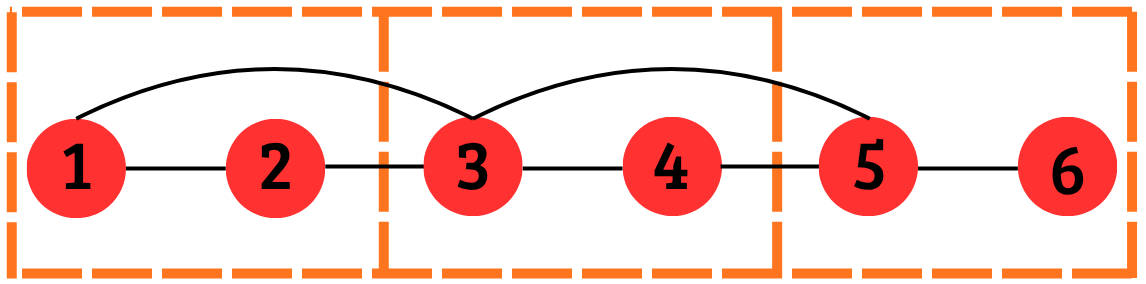}
\caption{Structure of the investigated delta chain with ferromagnetic nearest 
neighbor exchange $J_1$ and antiferromagnetic next-nearest neighbor exchange 
$J_2$ between adjacent odd sites.}
\label{thDMRGfMM-f-2}
\end{figure}

Several frustrated spin systems such as the delta (or sawtooth) chain 
feature unusual magnetic properties such as flat energy bands and
localized energy eigenstates, see e.g.\ \cite{KDN:PRB14,DmK:JPCM23}. 
These spectral properties arise for special ratios of the two
involved exchange interactions $J_1$ and $J_2$ in the Hamiltonian
\begin{align}
    \op{H}
    =& J_1 \sum_{i=1}^{N-1} 
    \vecopsprod{i}{i+1} 
    + J_2 \sum_{i=1}^{N/2-1}
    \vecopsprod{2i-1}{2i+1}
\label{ham2}
\\
&+ g \mu_B B \sum_{i=1}^{N} \ops{i}^z
\nonumber
    \ .
\end{align}
They may result in macroscopic magnetization jumps \cite{SHS:PRL02}
as well as an increased magnetocaloric effect \cite{ZhH:JSM04},
non-ergodic dynamics \cite{JES:PRB23}, and other frustration effects 
\cite{MBB:JPCM04,DRM:IJMPB15}.
For the version with ferromagnetic nearest neighbor exchange $J_1$ and 
antiferromagnetic next-nearest neighbor exchange $J_2$ between even neighbors, see
\figref{thDMRGfMM-f-2}, a quantum phase transition occurs for a critical ratio of
$\alpha=|J_2/J_1|$ \cite{KDN:PRB14}. For this case, chemical compounds exist  
\cite{INK:JPSJ05,BML:npjQM18} of which Fe$_{10}$Gd$_{10}$ happens to be close 
to the quantum critical point, compare \cite{BML:npjQM18}.

We repeat the earlier investigation \cite{HRS:JMMM19} 
of the ferromagnetic-antiferromagnetic delta chain
at $\alpha=0.55$ which for single spins $s_i=1/2$ is close to the
quantum phase transition ($\alpha_c=0.5$). 
For this scenario a low-energy scale appears
that gives rise to a low-temperature maximum of the specific heat 
which serves as a valuable feature to assess the accuracy of methods.
In the following, we will compare the matrix representation of thDMRG used
in \cite{HRS:JMMM19} to the two methods discussed in Sec.~\xref{sec-2}. 
FTLM is employed to detect the point of divergence.

\begin{figure}[ht!]
\centering
\includegraphics*[width=0.90\columnwidth]{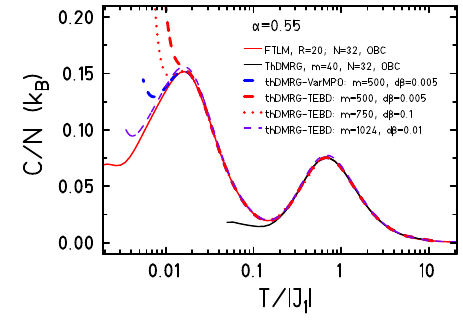}
\caption{Heat capacity for a delta chain: $N=32$, $\alpha=0.55$, open boundary condition. 
Both methods, thDMRG-TEBD and thDMRG-VarMPO are used and compared to FTLM
(red solid curve) as well as 
to an earlier computation (black solid curve) \cite{HRS:JMMM19}.
For details and computing times see text.}
\label{thDMRGfMM-f-3}
\end{figure}

Figure~\xref{thDMRGfMM-f-3} shows the specific heat for $N=32$ spins of $s_i=1/2$ 
at $\alpha=0.55$ calculated in four different ways. The FTLM result (red solid curve) 
is considered quasi-exact and serves as a benchmark. Our earlier investigation 
\cite{HRS:JMMM19} could not reproduce the double-peak structure, it was accurate
only down to temperatures $T/|J_1|\sim 0.1$ 
(black solid curve). 
The investigated new methods thDMRG-TEBD and thDMRG-VarMPO are both able to follow
the FTLM result down to the low-temperature maximum at $T/|J_1|\sim 0.01$ and
even beyond with reasonable choices of bond dimension $m$ and imaginary-time step
$d\beta$. Since the systematic dependence on the bond dimension $m$ and
the step $d\beta$ is not obvious, our advice for other users would be to run 
a few parameter sets in order to see which parts of the thermodynamic function
appear to be converged.

Computing times are determined with the same time-evolution parameters
of $m=500$, $d\beta=0.005$ and rules for increasing the bond dimension, 
on the same hardware  with the same number of cores. 
For these calculations VarMPO took 42 hours, while TEBD took 58 hours.
A major factor in the difference in computational time and performance 
can be explained by the grouping needed in TEBD. By grouping two sites 
the effective local spin has to cover the total dimension and thus
becomes $s = 3/2$ \cite{tenpy2024}. Adding the purification copy increases the 
effective spin to $s = 15/2$ and thus results in a chain of $L=16 , s=15/2$. 
In contrast, the VarMPO chain stays a $L=32 , s=3/2$.
In the end, both chains represent the same physical system and should yield 
the same results, however, choosing the same bond dimension $m= 500$ 
(between effective spins) does not result in the same accuracy and performance.
This simple example already shows how disruptive the grouping can become. 
While TEBD is in general the more robust and efficient method one should always consider 
if the increased computational cost is actually needed. 
However, since in this paper we focus mostly on open nearest-neighbour systems, 
i.e., less grouping, we will focus on TEBD.

\begin{figure}[ht!]
\centering
\includegraphics*[width=0.90\columnwidth]{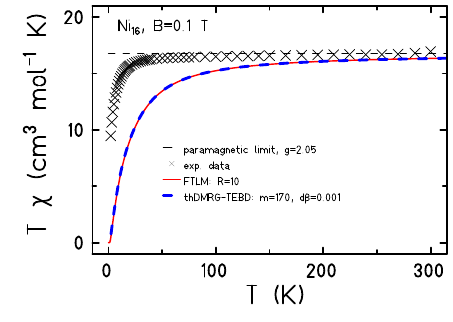}
\caption{Magnetic susceptibility $\chi=M/B$ times temperature $T$ at $B=0.1$~T 
for a nickel spin ring. Symbols show the data of \cite{DSR:CAAJ24},
the thin red curve shows the result obtained with FTLM,
and the blue dashed curve depicts the result
obtained with thDMRG-TEBD. 
$g=2.05$ is adapted from the experimental work.}
\label{thDMRGfMM-f-4}
\end{figure}

\subsection{A 16-membered Ni(II) ring}

In magneto-chemistry exchange interactions are often evaluated by means of
density functional theory (DFT). Experience taught us that these numbers 
often come with larger uncertainties. For small enough quantum spin systems 
an assessment and a refinement can be achieved by comparison to 
experimental observables via the evaluation of thermodynamics quantities. 
However, if the quantum spin system is too large 
for the employed method it remains unclear whether the DFT-values are 
appropriate.

Here, we demonstrate that both FTLM as well as thDMRG-TEBD are capable 
of evaluating the magnetic susceptibility for a spin ring of $N=16$ 
nickel spins $s_i=1$. Such a system was recently synthesized, and the exchange 
integrals were evaluated by means of DFT \cite{DSR:CAAJ24}. Our investigation
serves two purposes, we show that FTLM and thDMRG-TEBD yield the same 
result, and we show how this result compares to the experimental data,
see \figref{thDMRGfMM-f-4}. 
The  exchange interactions of four adjacent spin pairs were
estimated as $J=[28.08; -14.39; 7.22; 2.42]$~K repeating 
cyclically \cite{DSR:CAAJ24}. Our theoretical calculations demonstrate
that the interactions estimated by DFT are too strong.

\begin{figure}[ht!]
\centering
\includegraphics*[width=0.90\columnwidth]{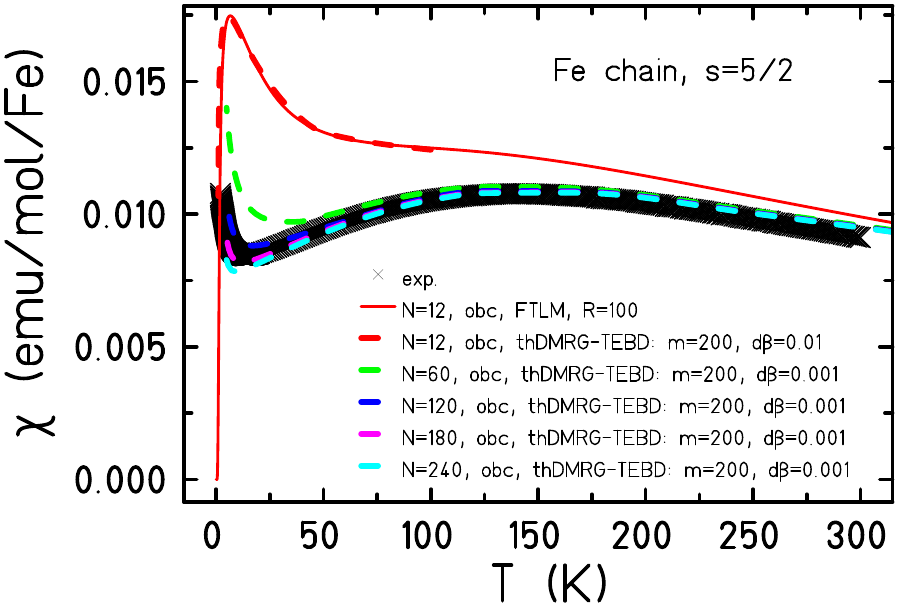}
\caption{Magnetic susceptibility $\chi=M/B$ at $B=0.1$~T for open chains. 
Experimental results (symbols) and the antiferromagnetic exchange interaction 
are taken from Ref.~\cite{WYH:ACIE}. 
thDMRG-TEBD calculations are given by dashed curves.
For the $N=12$ chain we did a calculation with bond dimension 
$m = 200$ and cooling steps $d\beta = 0.01$ up to $\beta_{max} = 2$. 
For the other chains calculation were done with  $m=200$ and $d\beta = 0.001$ up to $\beta_{max} = 0.2$  
for $N=\{60; 120; 180; 240 \}$.}
\label{thDMRGfMM-f-5}
\end{figure}

\subsection{Fe(III) chains and rings}

Molecular iron rings as well as chains appear often in quantum magnetism. 
The two examples which we discuss are an open iron chain recently investigated
in Ref.~\cite{WYH:ACIE} as well as a ten-membered ferric wheel 
\cite{TDP:JACS94,ScS:IRPC10}, both of iron(III) ions with $s_i=5/2$.
The Hamiltonian features nearest neighbor 
exchange either without or with periodic boundary conditions as well as a 
Zeeman term
\begin{align}
    \op{H}
    =& J \sum_{i=1}^{N} 
    \vecopsprod{i}{i+1} 
+ g \mu_B B \sum_{i=1}^{N} \ops{i}^z
\ .
\end{align}
Figure \xref{thDMRGfMM-f-5} shows the magnetic susceptibility $\chi=M/B$ at $B=0.1$~T 
for open chains of various length. FTLM can only reach $N=12$, whereas thDMRG
can deal with large system sizes since the open chain is perfectly suited 
for DMRG. This puts thDMRG into a favourable position since we can now address
very large chains. These show an upturn of the susceptibility at very low 
temperature in accordance with the experimental data.

We calibrated thDMRG with the FTLM-result for the open $N=12$ chain,
compare \figref{thDMRGfMM-f-5},
which aligns as intended, and thDMRG even catches the decrease of $\chi$ at 
rather low temperatures when using a large enough bond dimension of $m=200$. 
However, the susceptibility for $N=12$ is nowhere near the experimental data. 
Upon increasing the length of the chain the susceptibility
approaches the experimental data
and also develops an upturn at low temperature.

It remains, however, open whether the deviations of the low-temperature
behavior between the largest investigated chains are due to still insufficient 
convergence of thDMRG-TEBD or due to the experimental observation that
spin chains in real materials are interrupted by impurities and thus
constitute a distribution of various chain lengths \cite{EgA:JMMM04,MKW:PRB13}. 

\begin{figure}[ht!]
\centering
\includegraphics*[width=0.90\columnwidth]{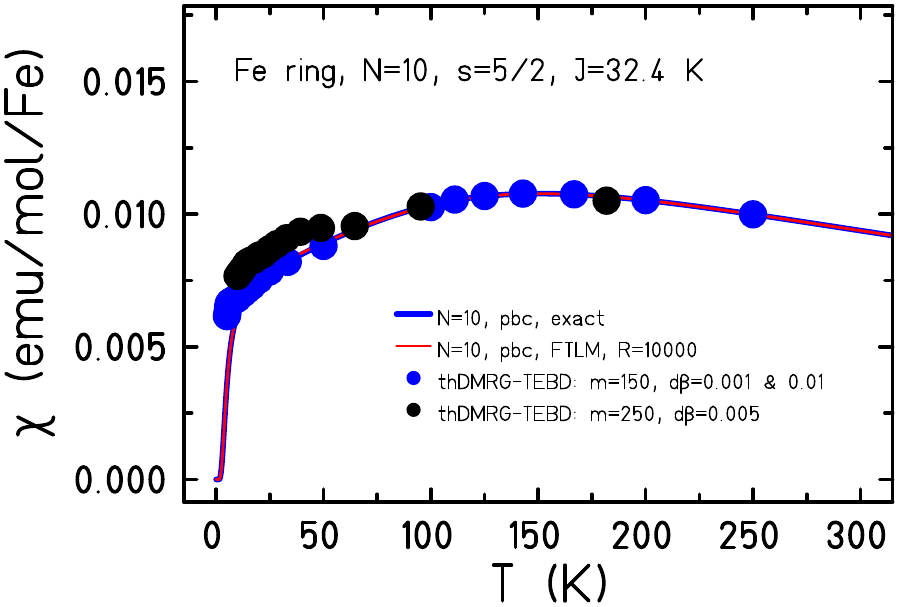}
\caption{Magnetic susceptibility at $\chi=M/B$ $B=0.1$~T for 
a spin ring of $N=10$ spins $s_i=5/2$ and again $J=32.4$~K.
Theoretical calculations by means of exact diagonalization 
\cite{ScS:IRPC10} and FTLM \cite{SRS:PRR20} are given by various curves.
Results of thDMRG-TEBD for two parameter sets are given by bullet points.}
\label{thDMRGfMM-f-6}
\end{figure}

The periodic variant behaves rather differently as can be seen in \figref{thDMRGfMM-f-6},
compare also \cite{KaS:E26}.
We employed the same exchange constant to make it comparable to \figref{thDMRGfMM-f-5}.
The susceptibility does not feature a low-temperature maximum or upturn.

Theory-wise, we now deal with case \figref{thDMRGfMM-f-1}(c) where a site
consists of two fused spins with $s=5/2$ each. This results in a Hilbert space per site
of dimension $6 \times 6 = 36$. The computational effort grows considerably, and the 
accuracy suffers when using affordable bond dimensions.
Interestingly, this case can still be evaluated by exact diagonalization employing
symmetries \cite{ScS:IRPC10}. Therefore, we compare the results of
exact diagonalization (blue), FTLM (red), and thDMRG-TEBD (bullet points) in  
\figref{thDMRGfMM-f-6}. While the exact and the FTLM results agree, 
the thDMRG result is accurate at high temperatures, but at low temperatures a divergence is noticed. 
The origin is mostly a too small bond dimension at low $T$. One also notices that 
a smaller integration step $d\beta$ is not automatically better as is known from
ordinary numerical integration. However, the combination of small steps
$d\beta=0.001$ down to 100~K and larger steps $d\beta=0.01$ below does a 
good job. Small steps at small $\beta$ are needed to approximate observables 
at larger temperatures. 
The results are qualitatively sufficient but took weeks on 64 cores.

\begin{figure}[ht!]
\centering
\includegraphics*[width=0.85\columnwidth]{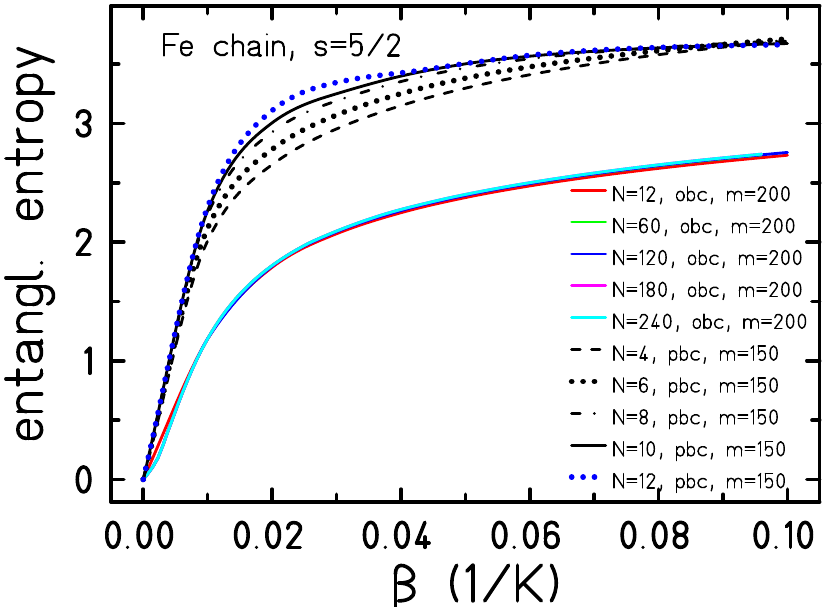}
\caption{Entanglement entropy for half the system as function of elapsed
imaginary time $\beta$.}
\label{thDMRGfMM-f-7}
\end{figure}

We would like to use this example to demonstrate that deviations from open
chains do have an impact on entanglement entropy, the area law and thus on 
numerical effort and achievable accuracy. Figure~\xref{thDMRGfMM-f-7}
shows the entanglement entropy for half the system (or a ratio that is close).
One notices that the entaglement entropy for the open iron chains (colored
curves) grows with $\beta$ which reflects the need for larger bond dimensions
in the course of imaginary time. However, the absolute values do not vary 
much with system size.

The entanglement entropy of the periodic chains (i.e. spin rings) is considerably
larger as well as more strongly dependent on the number of sites $N$
at least for smaller rings.

\subsection{Thermodynamics of a spin tube}

\begin{figure}[ht!]
\centering
\includegraphics*[width=0.99\columnwidth]{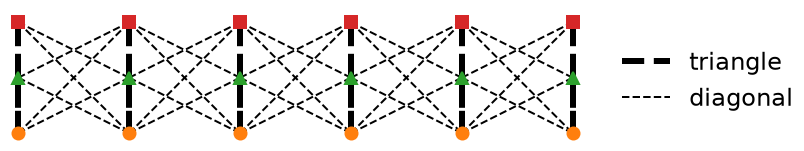}
\caption{Structure of the spin tube. Each triangle is an unitcell 
with nearest neighbour interactions and connected diagonal to the other two sites of the neighbour triangles.}
\label{thDMRGfMM-f-8}
\end{figure}

One of the most fascinating spin systems and a very rare structure as well
is given by a triangular spin tube made of copper ions, see \figref{thDMRGfMM-f-8}.
A first characterization by means of exact diagonalization of a periodic tube 
with six units in Ref.~\cite{SNK:PRB04} revealed that the two involved
exchange interactions $J_1=-0.9$~K within the triangles and $J_2=-1.95$~K between 
the triangles of the anti-prismatic tube are of rather similar magnitude. 
One would thus consider this tube to be highly frustrated. However, 
DMRG investigations of the ground state \cite{FLP:PRB06} showed 
that the tube is more similar to a Heisenberg chain of spins $s=3/2$
with just nearest-neighbor exchange at very low temperatures. 
This view could later be
confirmed by measurements of the specific heat \cite{ISS:PRL10}
which demonstrated a Tomonaga-Luttinger liquid behavior
corresponding to an effective spin-$3/2$ antiferromagnetic Heisenberg chain.

20 years after the first ground-state DMRG calculations we now try to model the
thermodynamic equilibrium properties by means of matrix-product states.
Interestingly, we faced a surprise.

\begin{figure}[ht!]
\centering
\includegraphics*[width=0.90\columnwidth]{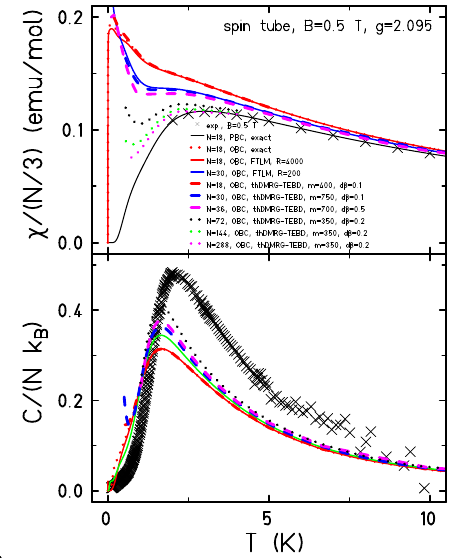}
\caption{Magnetic susceptibility $\chi=M/B$ per triangle at $B=0.5$~T (top) and
specific heat at $B=0$ (bottom) for a triangular spin tube. Symbols show the experimental  
results \cite{SNK:PRB04,ISS:PRL10}.
Theoretical results for exact calculations, FTLM, and thDMRG are displayed 
by various curves.}
\label{thDMRGfMM-f-9}
\end{figure}

As \figref{thDMRGfMM-f-9} shows there is a massive difference between 
spin tubes with periodic and open boundary conditions. 
The exchange interactions $J_1$ and $J_2$ were determined using
a periodic tube of only 18 spins, compare thin black curve in upper panel
of \figref{thDMRGfMM-f-9} and Ref.~\cite{SNK:PRB04}. 
This fits the susceptibility rather nicely. In retrospect, it would have been impossible
to determine the exchange interactions from calculations of small tubes
with open boundary conditions.

Our thDMRG approach employing TEBD for 18 spins and open boundary conditions
yields perfect agreement for the susceptibility
with the result of exact diagonalization for the same system,
even down to about one kelvin. The agreement with the largest possible
FTLM calculation of $N=30$ (using a normal workstation) is also very good good.
Employing thDMRG-TEBD we could access larger system sizes not possible
with FTLM to see
whether the theoretical susceptibility will convergence against 
the experimental data points. This seems to be the case, 
but happens rather slowly. It remains open, whether the compound would show
an upturn of the susceptibility at low temperatures 
since this was not measured at the time, unfortunately.

The calculations of the heat capacity demonstrate that thDMRG-TEBD agrees
very well with either exact diagonalization or FTLM. However, none of these
schemes can model the linear increase of the heat capacity at the lowest
temperatures characteristic of Luttinger liquids either since the system 
size is too small or low-enough temperatures cannot be reached by thDMRG.
For this, 100 effective spins $s=3/2$ in a linear effective chain 
had to be used in a quantum Monte-Carlo calculation in \cite{ISS:PRL10}.

\subsection{Breaking the ${S}^z$-symmetry in a $s=3/2$ chain}

One of our goals is to test the limits of the investigated thermal DMRG schemes. 
To this end, our final example deals with a spin system that features single-ion
anisotropy. This system is motivated by a recent investigation 
of a rhenium chain compound
that will be published elsewhere \cite{Arn:26}.

\begin{figure}[ht!]
\centering
\includegraphics*[width=0.99\columnwidth]{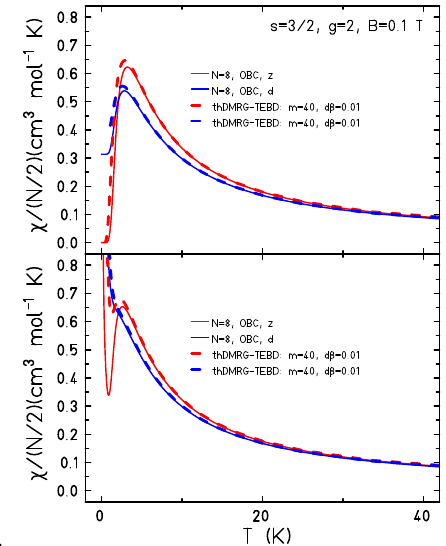}
\caption{Magnetic susceptibility $\chi=M/B$ per unit cell  
at $B=0.1$~T for collinear arrangements of easy axes (top) as well as
alternating canted arrangements (bottom).
$\chi$ is given along $z$-direction as well as powder average (label d). 
Results of exact diagonalization are depicted by solid curves 
and results of thDMRG-TEBD by dashed curves.}
\label{thDMRGfMM-f-10}
\end{figure}

The Hamiltonian of the system has the form 
\begin{align}
\label{ham3}
    \op{H}
    =& J \sum_{i=1}^{N} \vecopsprod{i}{i+1} 
    + D \sum_{i=1}^{N} \left(\vecops{i}\cdot \vec{e}_i \right)^2 
+ g \mu_B \vec{B}\cdot \sum_{i=1}^{N} \vecops{i}  
\end{align}
with $D<0$ being the anisotropy parameter. The spin quantum number is $s=3/2$.
We investigate two cases: a uniaxial arrangement where all $\vec{e}_i$ are along
$z$-direction and a canted case where every second $\vec{e}_i$ is canted by $45^\circ$
into the $x$-direction. For all calculations $J=1$~K and $D=-2$~K. As in typical 
experiments, we study the magnetization along $z$-direction as well as a 
powder average (over the directions given by the vertices of a dodecahedron).

From the point of symmetry, the new system lacks $\op{S}^z$-symmetry for two reasons:
(a) even if all easy axes are collinear, the external field may point somewhere else,
and (b) in the real compound the unit cell contains two directions of the easy axes.
For magnetic systems the loss of $\op{S}^z$-symmetry not only increases the dimensions
since the Hamiltonian looses its related block structure, also approximations
such as e.g.\ FTLM converge much more slowly \cite{HaS:EPJB14}.

However, for open anisotropic chains of length $N=8$ we observe a good agreement
as can be seen in \figref{thDMRGfMM-f-10}. All investigated cases are nicely
reproduced by thDMRG-TEBD for rather small bond dimensions of $m=40$. An exception 
is only the canted case where the magnetization is evaluated along the $z$-direction 
that coincides with every second easy axis.

Periodic boundary conditions would be simpler in case of exact diagonalization 
because of additional point group symmetries, but much more demanding for
thDMRG-TEBD since a unit cell now would have to contain four spins with a 
total dimension of 256 for the unit cell.

\section{Discussion}
\label{sec-4}

The present investigation rests on work by Hauschild and Pollmann  
\cite{HLB:PRB18,UHP:PRB23,tenpy2024},
in particular the TenPy numerical package (version 1.1.0). Our goal was to study
from the perspective of a user
whether this DMRG-based approach would enable us to approximate thermodynamic
equilibrium observables for quantum spin systems. In the present article,
we focus on quasi one-dimensional magnetic molecules and spin chains, however,
with spins larger than $1/2$, more than nearest neighbor interactions, and 
periodic as well as open boundary conditions.

Overall, our impression is very good. The method can address bipartite as well as
frustrated spin systems with and without $\op{S}^z$-symmetry for system sizes that
are larger than accessible by the finite-temperature Lanczos method.

As expected the numerical effort grows with more spins per unit cell
as is unavoidably given when addressing systems with periodic boundary conditions.
The resulting dimension of the Hilbert space of a unit cell puts a barrier
in front of larger calculations. We thus think that calculations for the
sawtooth ring molecule Fe$_{10}$Gd$_{10}$ \cite{BML:npjQM18} are out of reach
at the moment for thDMRG, but other tensor network 
methods like METTS \cite{SW:NJP10} or iPEPS and ODTNS as in \cite{CRL:SB18}
might be successful.

Another future investigation will address the accuracy at very low temperatures
as for instance done in Ref.~\cite{BiB:PRB15} where the approach discussed here is compared to METTS. 
This is a necessary prerequisite to rule out ambiguities connected to e.g.\ 
the upturn observed for the susceptibility at large $\beta$, i.e.\ small temperatures 
for large system sizes.
This not only requires larger computational resources as well as a systematic investigation
of the influence of bond dimension as well as imaginary-time step,
it also needs demanding complementary calculations e.g.\ using the Bethe ansatz
or Quantum Monte Carlo to compare to.

For the time being, our advice for users would be to vary the bond dimension $m$ as well as
the step $d\beta$ in order to see which parts of thermodynamic functions remain unaffected and
thus suggest to be converged.


\section*{Acknowledgment}

Calculations were performed using the TeNPy library (version 1.1.0)
\cite{tenpy2024}.
We thank Johannes Hauschild and Frank Pollmann (TU Munich) for fruitful
discussions about recent developments concerning approximate schemes 
building on matrix-product states as well as Frank Pollmann and Andreas Honecker 
for critical reading.

%
%

\end{document}